\documentclass[12pt]{iopart}

\usepackage{scalerel}[2014/03/10]
\usepackage{stackengine}
\usepackage{amsmath}
\usepackage{hyperref}
\usepackage{amsfonts}
\hypersetup{
    colorlinks=true,
    linkcolor=blue,
    filecolor=magenta,      
    urlcolor=cyan,
    pdftitle={Overleaf Example},
    pdfpagemode=FullScreen,
  }
  \usepackage{cite}

\newcommand{\taud}{\tau_d}

\newcommand{\mscriptsize}[1]{\mbox{\scriptsize\ensuremath{#1}}}
\newcommand{\mtiny}[1]{\mbox{\tiny\ensuremath{#1}}}
\newcommand{\diff}{\mscriptsize{\text{(p)}}}
\newcommand{\refr}{\mscriptsize{\text{(r)}}}
\newcommand{\free}{\mscriptsize{\text{(f)}}}
\newcommand{\opt}{\mscriptsize{\text{opt}}}
\newcommand{\optt}{\mtiny{\text{opt}}}
\newcommand{\inn}{\mscriptsize{\text{in}}}
\newcommand{\out}{\mscriptsize{\text{out}}}

\renewcommand{\d}{\mbox{d}}
\renewcommand\Tilde[1]{\ThisStyle{%
  \setbox0=\hbox{$\SavedStyle#1$}%
  \stackengine{-.1\LMpt}{$\SavedStyle#1$}{%
    \stretchto{\scaleto{\SavedStyle\mkern.2mu\sim}{.5467\wd0}}{.7\ht0}%
  }{O}{c}{F}{T}{S}%
}}

\begin{document}

\title[Stochastic resetting with refractory periods]{Stochastic resetting with refractory periods: pathway formulation and exact results}

\author{G. García-Valladares$^1$, D. Gupta$^{2,3}$, A. Prados$^1$, and C. A. Plata$^1$}
\address{$^1$Física Teórica, Universidad de Sevilla, Apartado de
  Correos 1065, E-41080 Sevilla, Spain}
\address{$^2$Department of Physics, Indian Institute of Technology Indore, Khandwa Road, Simrol, Indore-453552, India\\
$^3$Nordita,  Royal Institute of Technology and Stockholm University,  Hannes Alfvéns väg 12, 23, SE-106 91 Stockholm, Sweden}
\ead{ggvalladares@us.es, phydeepak.gupta@gmail.com, prados@us.es, cplata1@us.es}
\vspace{10pt}
\begin{indented}
\item[] March 2024
\end{indented}

\begin{abstract}
We look into the problem of stochastic resetting with refractory periods. The model dynamics comprises diffusive and motionless phases. The diffusive phase ends at random time instants, at which the system is reset to a given position---where the system remains at rest for a random time interval, termed the refractory period. A pathway formulation is introduced to derive exact analytical results for the relevant observables in a broad framework, with the resetting time and the refractory period following arbitrary distributions. For the paradigmatic case of Poissonian distributions of the resetting and refractory times, in general with different characteristic rates, closed-form expressions are obtained that successfully describe the relaxation to the steady state. Finally, we focus on the  single-target search problem, in which the survival probability and the mean first passage time to the target can be exactly computed. Therein, we also discuss optimal strategies, which show a non-trivial dependence on the refractory period. 
\end{abstract}

%
\vspace{2pc}

\section{Introduction}

Stochastic resetting \cite{Evans11_resetting,Evans11_optimal,Evans20_applications} has become a very prolific topic within the field of non-equilibrium statistical mechanics. Stochastic resetting or restart can be thought of as one of the most elementary examples of an intermittent search strategy \cite{Benichou11rmp,Oshanin_efficient_2009,Rojo_intermittent_2010,Chupeau_cover_2015}, simple enough to analytically address the study of many physical quantities of interest. On the one hand, it has been successfully used in many different applications, ranging from economics \cite{Stojkoski21_nonergodic,Stojkoski22_income,Stojkoski2022_autocorrelation,Santra2022_tax,Vinod2022_nonergodicity,Montero2022_discounting,Jolakoski23_income} to biochemical reactions \cite{Reuveni14_unbinding,Rotbart15_michaelis,Pal21_TUR,Reuveni16_fluctuations,Biswas23_rate} or ecology \cite{Roldan16_backtrack,Plata20_catastrophic,Pal20_home,Evans22_predator}, mostly motivated by the beneficial effect of restart for lowering the first passage time \cite{Evans11_resetting,Evans11_optimal,Evans13_noneq,Bhat16_deterministic,Ahmad19_vanishing,Bressloff20_directed,Faisant21_experiment,DeBruyne23_control}. On the other hand, it constitutes an excellent test bench for performing non-equilibrium research, providing comprehensive models to study non-equilibrium steady states (NESS) \cite{Majumdar15_Relaxation,Mendez16_stationary,Eule16_steady,Pal16_time,Evans18_tumble,Gupta19_underdamped,Basu19_exclusion,Pal19_invariants}, stochastic thermodynamics and fluctuation theorems \cite{Pal21_TUR,Fuchs16_thermodynamics,Pal17_integral,Busiello20_entropy,Gupta20_fluctuations,Gupta22_work}, large deviations \cite{Meylahn15_large,Harris17_large,denHollander19_additive,Monthus21_large,Smith22_large,Zamparo22_fluctuations}, or quantum restart \cite{Dhar15_quantum,Rose18_quantum,Mukherjee18_quantum,Wald21_quantum,Sevilla23_quantum,Dubey23_quantum}, to name just a few.

Originally, stochastic resetting was introduced as instantaneous events that restart a given natural dynamics without any respite \cite{Evans11_resetting,Evans11_optimal}. Nevertheless, this instantaneousness cannot represent a real physical situation, since the resets are cost-free---and any actual, physical, implementation thereof must involve some cost. This has led to investigate more refined and realistic models, where resetting ceases to be a costless operation for the system. Depending on the phenomena, different strategies---mainly, the insertion of new phases---have been proposed to tackle this flaw in the simplest resetting model. On the one hand, return phases that alternate with the natural dynamics have been introduced \cite{Pal19_invariants,Gupta22_work,Bodrova20_return,Gupta21_return,Gupta21_linear,Radice22_return, olsen2023thermodynamic}. On the other hand, a motionless phase \cite{Evans19_refractory,Maso19_residence}, termed as refractory period, may be introduced after the resetting, which can be envisioned as a recovery time payed after performing the reset. Even though the latter strategy still involves an instantaneous reset, models of instantaneous resetting with refractory period phases are physically sound approaches to describe certain chemical and biological reactions. A paradigmatic example is that of the action potential in neurons, where its firing is followed by a quiescent state, i.e. an ineffective time to any stimulus \cite{Fetz_1983,Maida16_Neuron}. 

In particular, the study of stochastic resetting with refractory periods has been shown to be useful in the context of enzymatic reactions following the Michaelis-Menten scheme \cite{Reuveni14_unbinding,Rotbart15_michaelis,Pal21_TUR,Reuveni16_fluctuations}.  Therein, an enzyme binds to a substrate in a reversible binding-unbinding reaction, which, in a second step from the bound state, release a certain product. Here, the unbinding step may facilitate the production of products, i.e. interruption of a task may improve its  accomplishment---which is the essence of optimal restart strategies. 

This work focuses on the detailed analysis of stochastic resetting with refractory periods. Specifically, we provide a pathway formulation based on the statistics of any possible reset history of the system. Such a formulation is related to renewal theory \cite{Evans18_tumble,Chechkin18_renewal,Bodrova19_scaled,Wang21_renewal}, being inspired by similar techniques in different resetting setups \cite{Gupta22_work,Chechkin18_renewal}. We prove the validity of our pathway approach in a very broad framework, which allows us to obtain general results for the case of stochastic resetting with refractory periods.\footnote{Our general framework reproduces the specific results derived for some particular cases already considered in the literature \cite{Evans19_refractory,Maso19_residence}.} Moreover, exact results for the case of Poissonian resets with Poissonian refractory periods are derived. For that relevant situation, the evolution of the probability density distribution (PDF) of a resetting Brownian particles in an infinite domain with refractory periods is explicitly worked out. Also, we obtain the mean first passage time (MFPT) as a function of the rates governing the exponential distributions for both dynamical phases. Interestingly, the minimization of the MFPT that we carry out reveals that the optimal restart rate depends on the typical duration of the refractory periods after the interruption.

The rest of the article is organised as follows. The fundamental ingredients of the model are described in section \ref{sec:model}. Section \ref{section:ResettingPathways} is devoted to the detailed analysis of the PDF of the system through our pathway formulation. We explicitly obtain the whole evolution of the system, which reaches a NESS in the long-time limit. Section  \ref{sec:MFPT} deals with the MFPT. In addition to reobtaining a general expression for the MFPT wih refractory periods within our framework, an explicit formula in the case of both Poissonian resets and refractory periods is derived. Additionally, the optimal resetting rate is obtained as a function of the rate governing the duration of the refractory periods. The conclusions of our work are summarised in section \ref{sec:concl}. Finally, extensions of our results and some technicalities are discussed in the appendices.

\section{Stochastic model}
\label{sec:model}

We consider a quite general stochastic resetting process. Let the system be represented by a particle that, in absence of resetting, stochastically propagates following a distribution $p^{\free}$---governed by a Fokker-Planck equation, which we call ``natural'' or ``propagation'' dynamics. On top of this natural dynamics, random resets to a certain position $x_r$ occur. Time events at which the particle instantaneously goes back to $x_r$ are named resetting events, and denoted by $t_i$, where the subscript $i=1,2,\ldots$, stands for the order of occurrence. The probability that a resetting event takes place in the time interval $(t,t+\d t)$, is  $\d t\,f(t)$, so the integral
\begin{equation}
    F(t)=\int_{t}^\infty \d t' \, f(t'),
\end{equation}
is the probability that no resetting events have occurred up to time $t$. In other words, $F(t)$ is the probability of having an uninterrupted propagation phase lasting $t$ at least.

\begin{figure}
    \centering
    \includegraphics[width=0.8\textwidth]{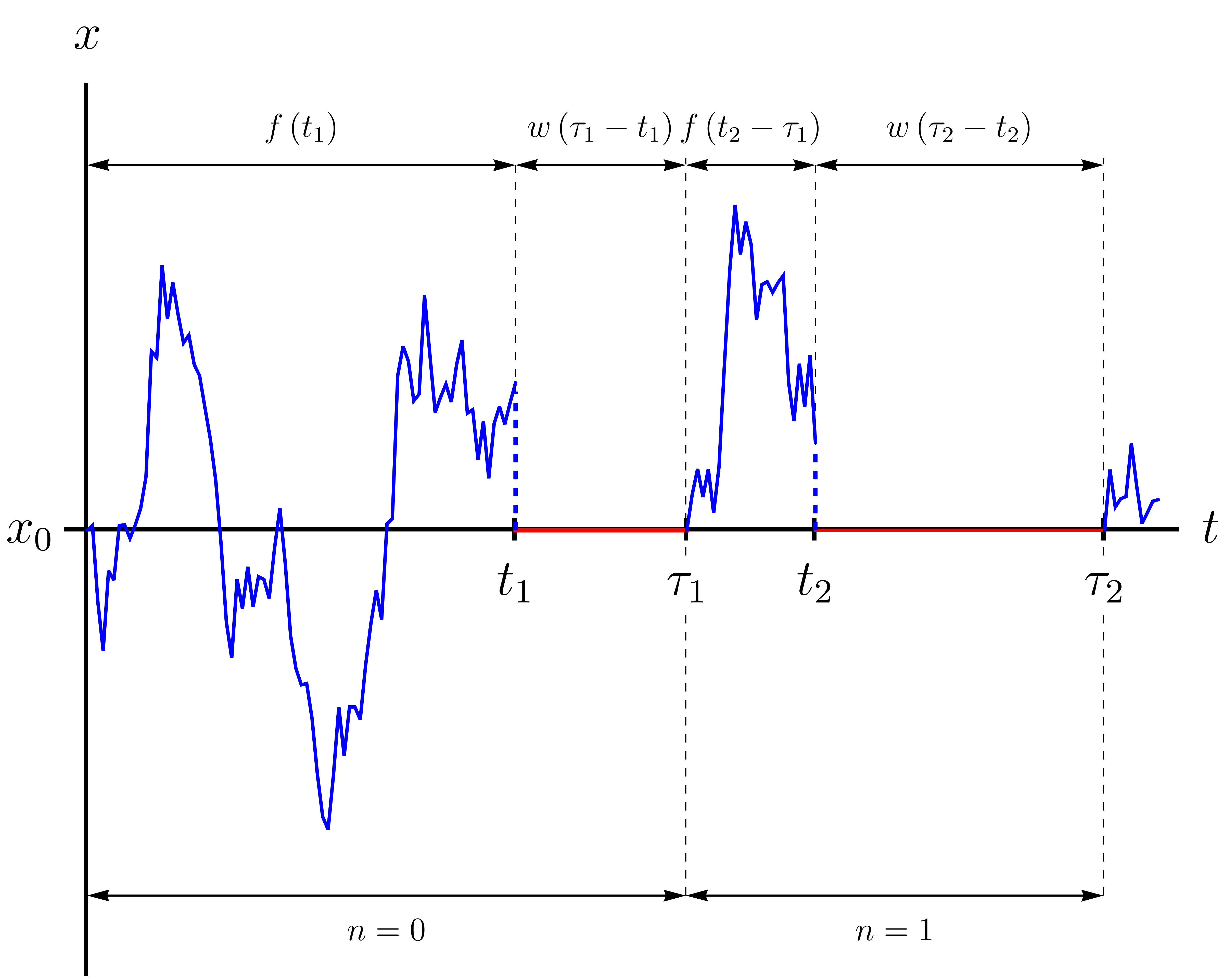}
    \caption{Single trajectory for stochastic resetting with refractory periods. The labels $t_i$, $i=1,2,\ldots$ , mark a new reset to $x_0$ where the propagation phase (blue line) ends with an instantaneous reset (dashed black line) and the refractory period begins (red line). Analogously, $\tau_i$ marks the end of the refractory phase after $t_i$, $i=1,2,\ldots$
    Herein, $n$ stands for the number of times the system is renewed, $n=i$ for $\tau_i < t < \tau_{i+1}$. The duration $\sigma$ of each propagation (refractory) phase comes from the PDF $f(\sigma)$ ($w(\sigma)$). For the sake of simplicity, the initial condition is equal to the resetting position $x_r=x_0$.}
    \label{fig:Model}
\end{figure}

In the simplest resetting process, the particle is instantaneously reset to  $x_r$ and carries on its natural dynamics---described by $p^{\free}$---right after. Instantaneous resets are difficult to motivate within the context of a realistic dynamics, since they involve an infinite energetic payment which is followed by no recuperation phase.  
With this problem in mind, we thoroughly analyse herein the effect of \emph{refractory periods}---random resting times after the reset \cite{Evans19_refractory,Maso19_residence}. Specifically, the particle is assumed to be at rest at $x_r$ after the $i$-th resetting event up to time $\tau_i$, for an independent random time $\sigma_i=\tau_i-t_i$. It is handy to introduce the number of renewals $n$ that the system has completed up to time $t$: specifically, $n=i$ when $\tau_{i}<t<\tau_{i+1}$, where $\tau_0=0$ is defined for consistency. The refractory period duration $\sigma$ is characterised by the PDF $w(\sigma)$, and the integral
\begin{equation}
    W(\sigma)=\int_{\sigma}^\infty \d \sigma' \, w(\sigma'),
\end{equation}
is the probability of having a refractory period longer than $\sigma$. In other words, $W(\sigma)$ is the probability of having a minimum refractory period equal to $\sigma$. An illustrative portrayal of this resetting dynamics for a one-dimensional model is shown in figure \ref{fig:Model}, where blue and red stand for the propagation and refractory phases, respectively.

In the following, we take the renewal structure of the resetting mechanism into account, where the first resetting event at $t_1$ and the first renewal at $\tau_1$ are considered. The PDF $p(x,t|x_0)$ of finding the particle in $x$, starting from $x_0$, after a time evolution of duration $t$ can be built as
\begin{eqnarray}\label{eq:p-x-t-x0}
    p(x,t|x_0)= &F(t)p^{\free}(x,t|x_0) \nonumber\\
    &+\int_0^t \d t_1 \, f(t_1) W(t-t_1) \delta (x-x_r) \nonumber\\
    &+\int_0^t \d t_1 \, f(t_1) \int_{t_1}^t \d \tau_1 \, w(\tau_1-t_1) \,p(x,t-\tau_1|x_r).
\end{eqnarray}
where $p^{\free}(x,t|x_0)$ is the free propagator of the natural dynamics in absence of resetting and $\delta(x)$ is the Dirac delta distribution. This is the renewal equation, which is an implicit equation for the PDF $p(x,t|x_0)$. Note that our writing for the renewal equation relies on the approach based on the first resetting, in opposition to approaches relying on the last resetting event, which are often used \cite{Evans20_applications}. The first term on the right hand side in equation \eref{eq:p-x-t-x0} is the contribution from trajectories where there has been no resetting up to time $t$, thus weighted by the probability $F(t)$. The second term stems from paths for which there has been a resetting event in the interval $(0,t)$ and the subsequent refractory phase has not ended at time $t$; therefore it contributes with $\delta(x-x_r)$. The last term comes from trajectories where the particle has been reset at $t_1\in(0,t)$, has had a subsequent refractory period finishing at $\tau_1\in(t_1,t)$, and has reached $x$ at time $t$ following its renewed dynamics in the interval $(\tau_1,t)$. For the sake of simplicity, we are going to take $x_r = x_0$ in the remainder of the paper, i.e.~the particle starts at the very beginning of the process from the resetting location.

\section{Pathway formulation for the probability density function}\label{section:ResettingPathways}

\subsection{General framework}

We aim at working out an explicit expression for $p(x,t|x_0)$ for all times, thus going beyond the solution of the PDF in the Laplace domain that can be found in the literature \cite{Evans19_refractory,Maso19_residence}. First, the density probability can be split into different pathways, based on how many renewals of the dynamics have occurred from the beginning,
\begin{eqnarray}
    p(x,t|x_0)=\sum_{n=0}^\infty \left[ p_n^{\diff}(x,t|x_0)+p_n^{\refr}(x,t|x_0) \right]. \label{eq:pn}
\end{eqnarray}
On the right-hand side (rhs) of equation \eref{eq:pn}, we have distinguished between the contribution of each phase---propagation ($\mbox{p}$) and refractory ($\mbox{r}$), depending on the final stage of the evolution at time $t$. Specifically, they are defined as follows: $p_n^{\mbox{\scriptsize{(s)}}}$ is the joint probability density function of observing the particle at position $x$ after exactly $n$ renewals of the dynamics, being $s\in\{\mbox{p},\mbox{r}\}$ the current phase at time $t$. 

The particular case $n=0$ corresponds to the no-renewed evolution,
\begin{equation}
    p_0^{\diff}(x,t|x_0)=F(t)p^{\free}(x,t|x_0), \quad p_0^{\refr}(x,t|x_0)= \delta(x-x_0) \int_{\tau_0}^{t} \d t_1 f(t_1-\tau_0) W(t-t_1),
    \label{eq:p0_def}
\end{equation}
where the system has not undergone the first resetting yet or has not finished the first refractory period, respectively. For generic $n \geq 0$, both $p_n^{\diff}$ and  $p_n^{\refr}$ can be built in a systematic way,
\begin{subequations}\label{eq:pn_general}
\begin{eqnarray}
    p_n^{\diff}(x,t|x_0)=& \prod_{i=1}^n \left[
         \int_{\tau_{i-1}}^{t} \d t_i\,f(t_i-\tau_{i-1})
        \int_{t_i}^{t} \d \tau_i\,w(\tau_i-t_i)
    \right] \nonumber\\
    &\times F(t-\tau_n)p^{\free} (x,t-\tau_n|x_0), \label{eq:pndiff_general}
\end{eqnarray}
\begin{eqnarray}
    p_n^{\refr}(x,t|x_0)=&\delta(x-x_0) \prod_{i=1}^{n} \left[
         \int_{\tau_{i-1}}^{t} \d t_i\,f(t_i-\tau_{i-1})
        \int_{t_i}^{t} \d \tau_i\,w(\tau_i-t_i)
    \right] \nonumber\\
    &\times \int_{\tau_{n}}^{t} \d t_{n+1}\,f(t_{n+1}-\tau_{n})
    W(t-t_{n+1}) , \label{eq:pnrefr_general}
\end{eqnarray}
\end{subequations}
where we recall that $\tau_0=0$ has already been introduced for the sake of a compact notation.\footnote{Note that \eqref{eq:pndiff_general} and \eqref{eq:pnrefr_general} admit a recursive relation between two consecutive renewals, specifically $p_n^{\mbox{\scriptsize{(s)}}}(x,t|x_0)=\int_{0}^{t} \d t_1\,f(t_1) \int_{t_1}^{t} \d \tau_1\,w(\tau_1-t_1)p_{n-1}^{\mbox{\scriptsize{(s)}}}(x,t-\tau_1|x_0)$.}

The construction of our solution is physically guided by the concept of resetting pathway: the total probability is the sum of the probabilities corresponding to each possible resetting pathway, weighted by the probability of that pathway.
Pathways that are in propagation and refractory period phases at time $t$ contribute with the conditional evolution during this last phase: the free propagator $p^{\free} (x,t-\tau_n|x_0)$ and the Dirac delta distribution $\delta(x-x_0)$, respectively. These contributions are weighted with the product of the having finished all the previous phases and having not finished the last one. Thus, our pathway formulation allows us to obtain an explicit solution of the renewal equation in the time domain, which constitutes an advantage with respect to previous approaches that rely on Laplace transforming the implicit renewal equation to find the solution in the Laplace domain.

The equations above can be simplified if we rewrite them as convolutions,
\begin{subequations}
\begin{eqnarray}
    p_n^{\diff}(x,t|x_0)=\left\{
        \left[ f*w\right]^{*n} *p_{0}^{\diff}\right\} (x,t|x_0),
        \label{eq:pndiff_convolution}
    \\
    p_n^{\refr}(x,t|x_0)=\delta(x-x_0)\left\{
        \left[ f*w\right]^{*n}*f*W
    \right\} (t)\label{eq:pnrefr_convolution},
\end{eqnarray}
\end{subequations}
where $p_0^{\diff}$ is defined in equation \eref{eq:p0_def}. Note that we have introduced the asterisk notation for the convolution product
\begin{equation}
[A * B] (t) = \int_0^t dt' A(t') B(t-t')
\end{equation}
and the convolution power $A^{*2}=A*A$.

Taking advantage of our expressing $p_n^{\diff}$ and $p_n^{\refr}$ as convolutions, their Laplace transforms are written in a straightforward way:
\begin{subequations}
\begin{eqnarray}
    \Tilde{p_n^{\diff}} (x,s|x_0)=\left(\Tilde{f} (s) \Tilde{w} (s)\right)^n \Tilde{p_{0}^{{\diff}}} (x,s|x_0), 
        \label{eq:pndiff_Laplace}
   \\    \Tilde{p_n^{\refr}}(x,s|x_0)=\left(\Tilde{f}(s)\Tilde{w}(s)\right)^{n}\Tilde{f}(s)\Tilde{W}(s)\delta(x-x_0).
    \label{eq:pnrefr_Laplace}
\end{eqnarray}
\end{subequations}
The sum over $n$ gives us the total contribution,
\begin{subequations}\label{eq:p_Laplace}
\begin{eqnarray}
    \Tilde{p^{\diff}} (x,s|x_0)=\sum_{n=0}^{\infty}\Tilde{p_n^{\diff}}(x,s|x_0)
    =\frac{\Tilde{p_{0}^{\diff}} (x,s|x_0)}{1-\Tilde{f}(s)\Tilde{w}(s)},
        \label{eq:pdiff_Laplace}
    \\
    \Tilde{p^{\refr}} (x,s|x_0)=\sum_{n=0}^{\infty}\Tilde{p_n^{\refr}}(x,s|x_0)
    =\frac{\Tilde{f}(s)\Tilde{W}(s)}{1-\Tilde{f}(s)\Tilde{w}(s)} \delta(x-x_0).
        \label{eq:prefr_Laplace}
\end{eqnarray}
\end{subequations}
This expression, obtained within our pathway formulation of the resetting process with refractory periods, was obtained in \cite{Maso19_residence} by a different approach.\footnote{This general expression reduces to equation (4) in \cite{Evans19_refractory} when diffusive propagation and Poissonian resetting are assumed.} Note that, in order to derive equation \eref{eq:p_Laplace}, we have considered that resetting and refractory periods are independent. However, our pathway framework can also be used to address the case in which resetting and refractory periods are correlated. Although we focus on the uncorrelated case, we show how to apply the introduced framework in the following subsection.

\subsubsection{Correlated PDF of resetting and refractory period}
\label{app:correlated}

Above, we have considered the time interval between resetting events and the refractory periods to be independent. Nevertheless, we may consider a more general situation, as done in \cite{Evans19_refractory}, in which resetting events and refractory periods are correlated.

With the just described perspective, we define $h(\Delta_1,\Delta_2)$ as the joint PDF of having a propagation phase of duration $\Delta_1$ and a subsequent refractory period of duration $\Delta_2$. Clearly, the marginal distributions correspond to those introduced before,
\begin{align}
    f(\Delta_1)&=\int_0^\infty \d \Delta_2 \, h(\Delta_1,\Delta_2), &
    w(\Delta_2)&=\int_0^\infty \d \Delta_1 \, h(\Delta_1,\Delta_2).
\end{align}
Now, the probabilities for the corresponding phases lasting a time interval shorter than $t$ can be rewritten as
\begin{align}
    F(t)&=\int_t^\infty \d \Delta_1 \, f(\Delta_1)=
    \int_t^\infty \d \Delta_1 \int_0^\infty \d \Delta_2 \,h(\Delta_1,\Delta_2),\\
    W(t)&=\int_t^\infty \d \Delta_2 \, w(\Delta_2)=
    \int_t^\infty \d \Delta_2 \int_0^\infty \d \Delta_1 \, h(\Delta_1,\Delta_2).
\end{align}

In the correlated case we are considering, the probability density of finding the particle at position $x$ at time $t$ follows a renewal equation
\begin{eqnarray}
    p(x,t|x_0)= &F(t)p^{\free}(x,t|x_0) \nonumber\\
    &+\delta (x-x_r)\int_0^t \d t_1 \int_t^\infty \d \tau_1\, h(t_1,\tau_1-t_1)  \nonumber\\
    &+\int_0^t \d t_1 \int_{t_1}^t \d \tau_1 \, h(t_1,\tau_1-t_1) \,p(x,t-\tau_1|x_r),    
\end{eqnarray}
which we underline that is an implicit equation for the PDF $p(x,t|x_0)$. Analogously to the uncorrelated case, here we apply our pathway formulation to explicitly compute $p(x,t|x_0)$. Therefore, $p(x,t|x_{0})$ is split into the sum \eref{eq:pn}, with the $n$-th order propagation and refractory contributions given now by
\begin{subequations}
\begin{eqnarray}
    p_n^{\diff}(x,t|x_0)=& \prod_{i=1}^n \left[
         \int_{\tau_{i-1}}^{t} \d t_i
        \int_{t_i}^{t} \d \tau_i\,h(t_i-\tau_{i-1},\tau_i-t_i)
    \right] \nonumber\\
    &\times F(t-\tau_n)p^{\free} (x,t-\tau_n|x_0), \\
    p_n^{\refr}(x,t|x_0)=&\delta(x-x_0) \prod_{i=1}^{n} \left[
         \int_{\tau_{i-1}}^{t} \d t_i
        \int_{t_i}^{t} \d \tau_i\, h(t_i-\tau_{i-1},\tau_i-t_i)
    \right] \nonumber\\
    &\times \int_{\tau_{n}}^{t} \d t_{n+1} \int_t^\infty \d \tau_{n+1} h(t_{n+1}-\tau_{n},\tau_{n+1}-t_{n+1}), \label{eq:Appendix_pref_x_t}
\end{eqnarray}
\end{subequations}
which provide explicit expressions in the time domain, with no need of resorting to Laplace transformation.

In this case, we do not have a clear convolution structure because of the correlation. Notwithstanding, it is possible to simplify these expressions going to the Laplace domain, as shown below. We do so in order to compare with previous results \cite{Evans19_refractory}. Let us start with the propagation phase, 
\begin{eqnarray}
    \Tilde{p_n^{\diff}}(x,s|x_0)&=&\int_0^\infty \d t\,e^{-st} \prod_{i=1}^n \left[
         \int_{\tau_{i-1}}^{t} \d t_i
        \int_{t_i}^{t} \d \tau_i\,h(t_i-\tau_{i-1},\tau_i-t_i)
    \right] \nonumber\\
    &&\times F(t-\tau_n)p^{\free} (x,t-\tau_n|x_0), 
    \nonumber\\
    &=& \prod_{i=1}^n \left[
         \int_{\tau_{i-1}}^{\infty} \d t_i
        \int_{t_i}^{\infty} \d \tau_i\,h(t_i-\tau_{i-1},\tau_i-t_i)
    \right]
    \nonumber\\
    &&\times \int_{\tau_n}^\infty \d t \, e^{-st} F(t-\tau_n)p^{\free} (x,t-\tau_n|x_0).
\end{eqnarray}
Introducing a new integration variable $t'=t-\tau_n$, the last integral turns to
\begin{equation}
    \int_{0}^\infty \d t' \, e^{-s(t'+\tau_{n)}} F(t')p^{\free} (x,t'|x_0)=e^{-s\tau_n} \Tilde{F p^{\free}}(x,s).
\end{equation}
Afterwards, introducing $\tau_n'=\tau_n-t_n$ and $t_n'=t_n-\tau_{n-1}$, 
\begin{equation}
    \int_{\tau_{n-1}}^{\infty} \d t_n
    \int_{t_n}^{\infty} \d \tau_n\,h(t_n-\tau_{n-1},\tau_n-t_n)e^{-s\tau}  
    \Tilde{F p^{\free}}(x,s)
    =e^{-s\tau_{n-1}}\Tilde{h}(s,s)\Tilde{F p^{\free}}(x,s),
\end{equation}
where we have employed the notation
\begin{equation}
    \Tilde{h}(s,m)=\int_0^\infty \d t \,e^{-st}
    \int_0^\infty \d \tau \,e^{-m\tau} h(t,\tau).
\end{equation}
for the bivariate Laplace transform of the joint probability $h(t,\tau)$. Note that $\Tilde{f}(s)=\Tilde{h}(s,0)$ and $\Tilde{w}(s)=\Tilde{h}(0,s)$.
Iterating this procedure $n$ times, we obtain
\begin{equation}
    \Tilde{p_n^{\diff}}(x,s)=\Tilde{h}^n(s,s)\Tilde{F p^{\free}}(x,s),
\end{equation}
and the Laplace transform of $p^{\diff}(x,t)$ is given by
\begin{equation}
    \Tilde{p^{\diff}}(x,s)=\frac{1}{1-\Tilde{h}(s,s)}\Tilde{F p^{\free}}(x,s).
\end{equation}

A similar procedure can be carried out for the refractory contribution. The Laplace transform of equation \eref{eq:Appendix_pref_x_t} is
\begin{align}
    \Tilde{p_n^{\refr}}(x,s|x_0)=&\delta(x-x_0)
    \int_0^\infty \d t e^{-st}
    \prod_{i=1}^{n} \left[
         \int_{\tau_{i-1}}^{t} \d t_i
        \int_{t_i}^{t} \d \tau_i\, h(t_i-\tau_{i-1},\tau_i-t_i)
    \right] \nonumber\\
    &\times \int_{\tau_{n}}^{t} \d t_{n+1} \int_t^\infty \d \tau_{n+1} h(t_{n+1}-\tau_{n},\tau_{n+1}-t_{n+1})
    \nonumber \\
    =& \delta(x-x_0) \prod_{i=1}^{n+1} \left[
         \int_{\tau_{i-1}}^{\infty} \d t_i
        \int_{t_i}^{\infty} \d \tau_i\, h(t_i-\tau_{i-1},\tau_i-t_i)
    \right] \int_{t_{n+1}}^{\tau_{n+1}} \d t \, e^{-st}
    \nonumber \\
    =& \delta(x-x_0) \prod_{i=1}^{n+1} 
 \left[
        \int_{\tau_{i-1}}^{\infty} \d t_i
        \int_{t_i}^{\infty} \d \tau_i\, h(t_i-\tau_{i-1},\tau_i-t_i)
    \right]  \frac{e^{-s t_{n+1}}-e^{-s \tau_{n+1}}}{s}. \nonumber \\ 
    & \,
\end{align}
Integrating in an iterative way as before, one gets
\begin{align}
    \Tilde{p_n^{\refr}}(x,s)
    &=\delta(x-x_0)
    \frac{\Tilde{h}(s,0)-\Tilde{h}(s,s)}{s}\Tilde{h}^{n}(s,s) \implies 
    \Tilde{p^{\refr}}(x,s)
    =\delta(x-x_0)\frac{\Tilde{h}(s,0)-\Tilde{h}(s,s)}{s\left[1-\Tilde{h}(s,s)\right]}.
\end{align}
Taking both contributions into account, the Laplace transform of the total PDF $p(x,t)$ results to be
\begin{equation}
    \Tilde{p}(x,s)=\frac{1}{1-\Tilde{h}(s,s)}
    \left[
    \Tilde{F p^{\free}}(x,s) + \delta(x-x_0)\frac{\Tilde{h}(s,0)-\Tilde{h}(s,s)}{s}
    \right],
\end{equation}
which exactly agrees with equation (26) in ~\cite{Evans19_refractory}. This alternative derivation illustrates that our pathway formalism provides a powerful tool to address the study of physical quantities of interest in intermittent dynamics.

\subsection{Poissonian resetting and refractory period}
\label{sec:Pois2}

Let us consider now that the resettings events and the refractory periods both follow exponential distributions, but with a different rate:
\begin{subequations}
\begin{eqnarray}
    f(t)&=r_1 e^{-r_1 t} \implies F(t)&=e^{-r_1 t},\label{eq:Distribution_Resetting} \\
    w(t)&=r_2 e^{-r_2 t} \implies W(t)&=e^{-r_2 t}. \label{eq:Distribution_Refractory}
\end{eqnarray}
\end{subequations}
The Laplace transform of $f(t)$ and its integral are, respectively, $\Tilde{f}(s)=r_1/(r_1+s)$ and $\Tilde{F}(s)=1/(r_1+s)$, with analogous expressions for $\Tilde{w}$ and $\Tilde{W}$ with the change $r_1\to r_2$.

With these choices, the Laplace transforms of $p^{\diff}$ and $p^{\refr}$ in equation \eref{eq:p_Laplace} 
turn out to be
\begin{subequations}\label{eq:p_Laplace_Poissonian}
\begin{eqnarray}
    \Tilde{p^{\diff}} (x,s|x_0)&=
    \Tilde{p^{\free}}(x,s+r_1|x_0)
    \nonumber\\
    &\quad
    +\frac{r_1 r_2}{r_1+r_2}\left(\frac{1}{s}-\frac{1}{s+r_1+r_2}\right)\Tilde{p^{\free}}(x,s+r_1|x_0), 
    \label{eq:pdiff_Laplace_Poissonian}
    \\
    \Tilde{p^{\refr}} (x,s|x_0)&=
    \frac{r_1}{r_1+r_2}\left(\frac{1}{s}-\frac{1}{s+r_1+r_2}\right)\delta(x-x_0),
    \label{eq:prefr_Laplace_Poissonian}
\end{eqnarray}
\end{subequations}
which can be readily inverted,
\begin{subequations}\label{eq:p_final}
\begin{eqnarray}
    p^{\diff} (x,t|x_0)&=&
    e^{-r_1 t}p^{\free}(x,t|x_0)
    \nonumber\\
    &&+\frac{r_2}{r_1+r_2}r_1 e^{-r_1 t}\int_0^t \,\d \tau\,
    \left(e^{r_1 \tau}-e^{-r_2 \tau}\right)p^{\free}(x,t-\tau|x_0),
    \label{eq:pdiff_final}
    \\
    p^{\refr} (x,t|x_0)&=&
    \frac{r_1}{r_1+r_2}\left(1-e^{-(r_1+r_2)t}\right)\delta(x-x_0).
    \label{eq:prefr_final}
\end{eqnarray}
\end{subequations}
Therefore, we have obtained the exact evolution of the system in the time domain---although still in terms of an integral term. Later on, specifically in section \ref{sec:relax}, an approximation leading to a closed-form expression is provided. Note that the ``standard'' expressions---i.e. those corresponding to the case without refractory period---are reobtained by taking the limit $r_2\to\infty$  in equation \eref{eq:p_final}.

\subsubsection{The case \texorpdfstring{$r_1=r_2$}.}

When the Poissonian rates for resetting and refractory periods are equal, $r_1=r_2=r$, the above expressions become especially simple,
\begin{subequations}\label{eq:req}
\begin{eqnarray}
    p^{\diff} (x,t|x_0)&=
    e^{-r t}p^{\free}(x,t|x_0)+ r e^{-r t}\int_0^t \d \tau
    \sinh\left(r\tau\right)p^{\free}(x,t-\tau|x_0), \label{eq:prop_req}
    \\
    p^{\refr} (x,t|x_0)&=
    e^{-rt}\sinh\left(r t\right)\delta(x-x_0). \label{eq:refr_req}
\end{eqnarray}
\end{subequations}
We would like to highlight that these expressions may be directly obtained in the time domain, without resorting to the Laplace transform. Let us go back to equation \eref{eq:pn_general} and substitute the exponential distributions therein,
\begin{subequations}
\begin{eqnarray}
    p_0^{\diff} (x,t|x_0)
    &= e^{-rt} p^{\free}(x,t|x_0)
    \\ p_n^{\diff} (x,t|x_0)
    &=r^{2n}e^{-rt}\int_0^t \d t_1 \int_{t_1}^t \d\tau_1 \int_{\tau_1}^t \d t_2 \ldots
    \int_{t_n}^t \d \tau_n\,  p^{\free}(x,t-\tau_n|x_0), \nonumber\\
    &=r^{2n}e^{-rt}\int_0^t d\tau_n \frac{\tau_n^{2n-1}}{(2n-1)!}p^{\free}(x,t-\tau_n|x_0),  \quad n \geq 1,
\end{eqnarray}
\begin{eqnarray}
    p_n^{\refr} (x,t|x_0)
    =e^{-rt} \frac{(rt)^{2n+1}}{(2n+1)!}\delta(x-x_0).
\end{eqnarray}
\end{subequations}
Summing over all $n$ yields equation \eref{eq:req}.

\subsubsection{Non-equilibrium steady state.}
\label{section:DiffusiveProcess}

Now, the asymptotic long-time behaviour is derived. For doing so, we assume a specific functional form of the free propagator $p^{\free}(x,t|x_0)$. Specifically, we consider the most usual case, which is pure diffusion. Therein, $p^{\free}$ is the Green function for the diffusion equation,
\begin{equation}
p^{\free}(x,t|x_0)=\frac{1}{\sqrt{4\pi D t}}\exp\left[-\frac{(x-x_0)^2}{4Dt}\right], 
\label{eq:pure_difussion}
\end{equation}
with $D$ being the diffusion coefficient.
 
\begin{figure}
    \centering
    \includegraphics[width=0.8\textwidth]{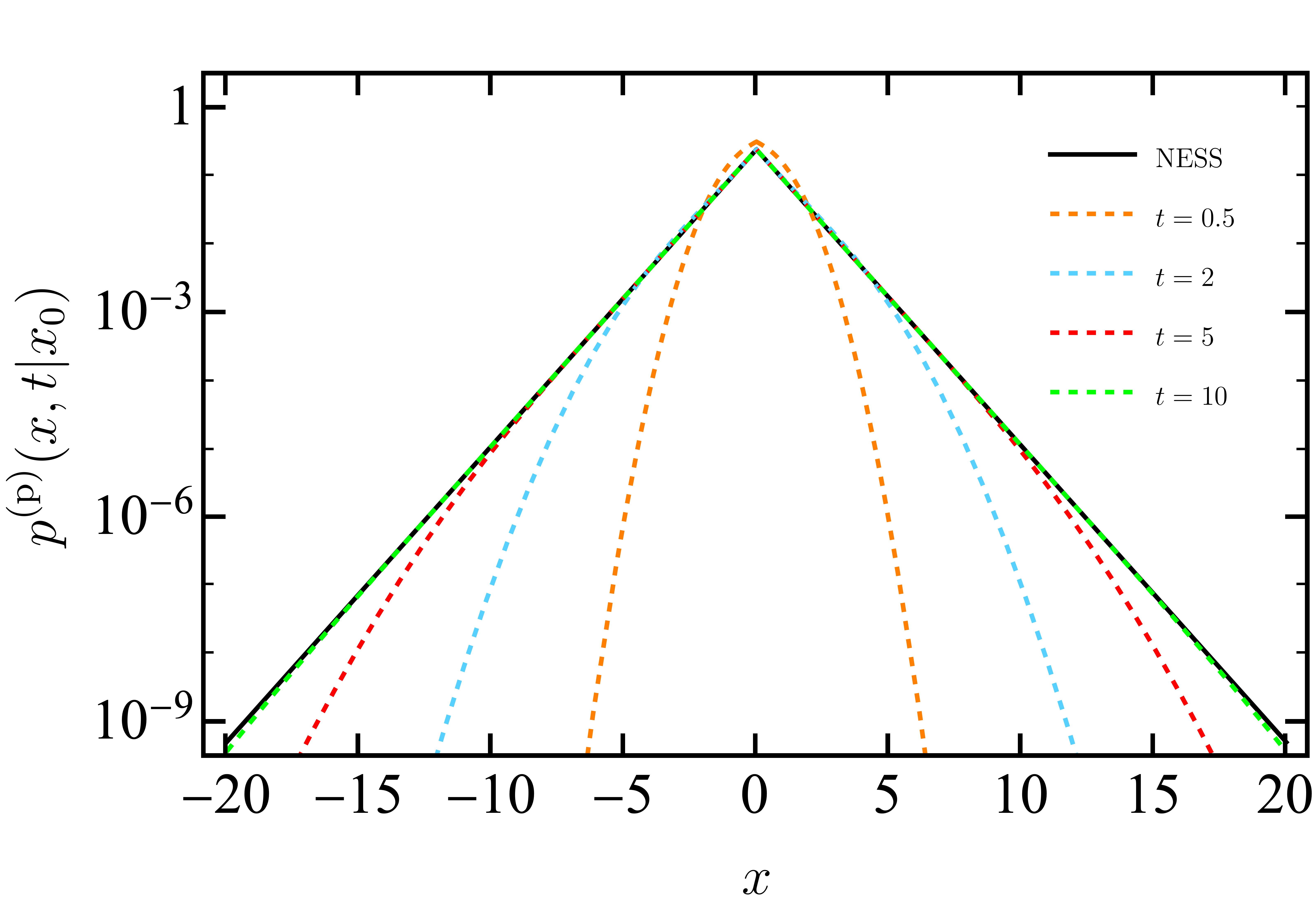} 
   \caption{PDF of the propagation phase. Numerical integration of $p^{\diff}(x,t|x_0)$ (colourful dashed lines), given by equation \eref{eq:pdiff_final}, at different times and the infinite time NESS \eref{eq:Ness_pprop} (solid black line) are shown. All the results are shown using $x_0=0$, $D=1$, $r_1=r_2=1$ as parameters.}
    \label{fig:Ness_NumericalIntegration}
\end{figure}

The long-time behaviour of $p(x,t|x_0)$  can be found by making use of the final value theorem
\begin{equation}
    \lim_{t\to\infty} z(t)= \lim_{s\to 0} s \,\Tilde{z}(s).
\end{equation}
Hence, taking into account equation \eref{eq:p_Laplace_Poissonian}, the long-time behaviour for arbitrary $(r_1, r_2)$ is achieved,
\begin{subequations}\label{eq:NESS}
\begin{align}
    &\lim_{t\to\infty} p^{\diff}(x,t|x_0)=
    \lim_{s\to 0} s \, \Tilde{p^{\diff}}(x,s|x_0) = 
    \frac{1}{2}\frac{r_2}{r_1+r_2}\sqrt{\frac{r_1}{D}}\exp\left[-\sqrt{\frac{r_1}{D}}|x-x_0|\right], \label{eq:Ness_pprop}
    \\
    &\lim_{t\to\infty} p^{\refr}(x,t|x_0)=
    \lim_{s\to 0} s \, \Tilde{p^{\refr}}(x,s|x_0) = 
    \frac{r_1}{r_1+r_2}\delta(x-x_0). \label{eq:NESS_prefr}
\end{align}
\end{subequations}
Of course, these results are consistent with those obtained by taking the infinite time limit in equation \eref{eq:p_final}, as well as with the results found in \cite{Evans19_refractory,Maso19_residence}. Note that the normalization of propagation and refractory phases in the stationary are given by the fraction of the average time spent in the corresponding phase, as physically expected. In figure \ref{fig:Ness_NumericalIntegration}, the convergence of the integral expression \eref{eq:pdiff_final}  of $p^{{\diff}}$ to its NESS \eref{eq:Ness_pprop} is shown.

\subsubsection{Relaxation to the steady state.}
\label{sec:relax}

We have already obtained exact expressions for the PDFs of each phase in the time domain, equations \eref{eq:pdiff_final} and \eref{eq:prefr_final}, as well as their long-time behaviour, equations \eref{eq:Ness_pprop} and \eref{eq:NESS_prefr}. Still, equation \eref{eq:pdiff_final} is not particularly illuminating, since one cannot infer how the relaxation to the NESS occurs in the propagation phase in a transparent way. Figure \ref{fig:Ness_NumericalIntegration} provides a hint on this issue, it can be observed that $p^{\diff}(x,t|x_0)$ reaches the steady state gradually in a central region $|x-x_0| < \hat{x}(t)$. The typical length around the reset point in which the NESS has been already reached, $\hat{x}(t)$, grows as $t$ increases, establishing a dynamic separation between a transient outer region, $|x-x_0| > \hat{x}(t)$, and the aforementioned relaxed inner region, $|x-x_0| < \hat{x}(t)$. Similar phenomena have already been observed in other resetting setups \cite{Majumdar15_Relaxation,Gupta19_underdamped}.

In order to obtain the long-time behaviour of equation \eref{eq:pdiff_final}, we start by rewriting  it as
\begin{eqnarray}
    \fl
    p^{\diff} (x,t|x_0)=&
    \frac{1}{\sqrt{4 \pi D t}} \exp \left[-t  \, \Phi_1 \left(1;\frac{x-x_0}{t} \right) \right]
    \nonumber\\
    &+\frac{r_1 r_2}{r_1+r_2} \sqrt{\frac{t}{4\pi D }}
    \int_0^1 \,\frac{\d \omega}{\sqrt{\omega}}\,\exp \left[-t \, \Phi_1\left(\omega;\frac{x-x_0}{t}\right) \right]
    \nonumber \\
        &-\frac{r_1 r_2}{r_1+r_2} \sqrt{\frac{t}{4\pi D }}
    e^{-(r_1+r_2)t} \int_0^1 \,\frac{\d \omega}{\sqrt{\omega}}\,\exp \left[ t \, \Phi_2\left(\omega;\frac{x-x_0}{t}\right) \right]  \label{eq:preLap}
\end{eqnarray}
where we have introduced the change of variable $\omega = 1 - \tau/t$, and defined
\begin{equation}
    \, \Phi_1(\omega;y)\equiv r_1\omega + \frac{y^2}{4 D \omega}, 
    \quad \, \Phi_2(\omega;y)\equiv r_2\omega-\frac{y^2}{4 D \omega}.
\end{equation}
Note that we assume pure diffusion, as stated in equation~\eqref{eq:pure_difussion}, for the sake of simplicity. For long times, taking constant $(x-x_0)/t$, the main contribution to the integrals over $\omega$ stem from the maximum of the exponents, i.e. the  minimum for $ \Phi_1(\omega;(x-x_0)/t)$ and the maximum for $ \Phi_2(\omega;(x-x_0)/t)$---as given by the so-called Laplace method for the asymptotic evaluation of integrals \cite{ByO99}. In the following, we give a simplified picture of the derivation of the dominant behaviour of equation \eref{eq:pdiff_final}---or \eref{eq:preLap}---stemming from the Laplace method, emphasising the intuitive ideas. A rigorous derivation of a more complex, but still explicit, formula for the long-time behaviour, which properly takes into account all the terms involved in equation \eref{eq:preLap}, as well as the subtleties stemming from the correct application of Laplace's method when the maximum is close to the boundaries of the integration interval, is relegated to \ref{app:Laplace-method}.

For long times, taking constant $(x-x_0)/t$, the main contribution to the integrals over $\omega$ in equation \eref{eq:preLap}  arises from the maximum of the exponents, i.e. the  minimum for $ \Phi_1(\omega;(x-x_0)/t)$ and the maximum for $ \Phi_2(\omega;(x-x_0)/t)$. On the one hand, $\Phi_2$ is a monotonically increasing function of $\omega$ and its maximum is always at the upper limit of integration. The corresponding contribution is thus always subdominant against the first, non-integral, term in equation \eref{eq:preLap}, i.e. the one involving trajectories with no resetting events, as shown in \ref{app:Laplace-method}. On the other hand, $\Phi_1(\omega;y)$ is not monotonic and has a single absolute  minimum at $\omega_0=|y|/\sqrt{4 D r_1}$, since $r_1$ and $D$ are strictly positive. The minimum of $\Phi_{1}$ within the integration interval $(0,1)$ is $\omega_{0}$ if $\omega_{0}<1$, and direct application of the Laplace method gives
\begin{equation}
    \frac{r_1 r_2}{r_1+r_2} \sqrt{\frac{t}{4\pi D }}\int_0^1 \,\frac{\d \omega}{\sqrt{\omega}}\, \exp \left[ -t \, \Phi_1\left(\omega;\frac{x-x_0}{t}\right)\right]
    \sim    \frac{1}{2}\frac{r_2}{r_1+r_2}\sqrt{\frac{r_1}{D}}\exp\left[-\sqrt{\frac{r_1}{D}}|x-x_0|\right],
\end{equation}
which corresponds to the NESS \eref{eq:Ness_pprop}. However, if $\omega_0 > 1$, the minimum of $\Phi_{1}$ within the integration interval is reached at the boundaries. Similarly to the situation with $\Phi_{2}$, this entails that the corresponding contribution is subdominant against the first term in equation \eref{eq:preLap}.

Summing up, the PDF of the propagation phase can be roughly estimated as 
\begin{equation}
    \fl
    p^{\diff} (x,t|x_0)  \sim
    \left\{ 
    \begin{aligned}        
        \frac{1}{2}\frac{r_2}{r_1+r_2}\sqrt{\frac{r_1}{D}}\exp\left[-\sqrt{\frac{r_1}{D}}|x-x_0|\right], \quad \frac{|x-x_0|}{t} < \sqrt{4 D r_1},\\
        \frac{1}{\sqrt{4 \pi D t}}\exp\left[-r_1 t -\frac{(x-x_0)^2}{4 D t}\right], \quad \frac{|x-x_0|}{t} > \sqrt{4 D r_1}.
    \end{aligned} \right. \label{eq:Approximation_NESS}
\end{equation}
The above discussion explains the observed separation into two regimes: the system has relaxed to the NESS within a certain spatial region, which increases linearly with $t$, as given by the condition $|x-x_0|/t < \sqrt{4 D r_1}$, whereas the transient behaviour is observed outside, i.e. for $|x-x_0|/t > \sqrt{4 D r_1}$.

\begin{figure}
    \centering
    \includegraphics[width=0.8\textwidth]{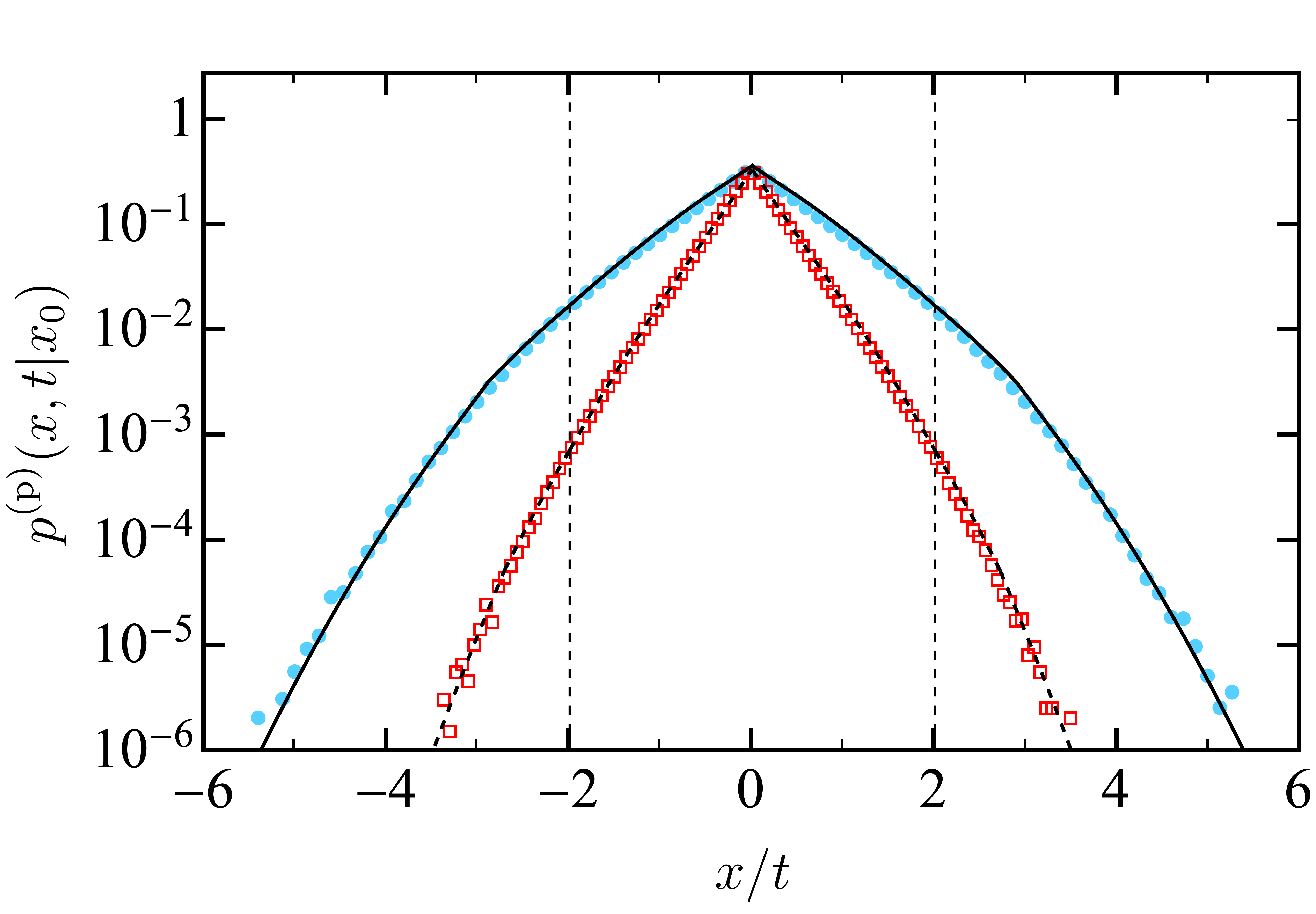}
    \caption{Comparison of the numerical and analytical PDFs of the propagation phase. Parameter values are $D=1$, $r_1=1$, $r_2=2$ and $x_0=0$.  Symbols stand for numerical simulations for $t=1.5$ (light blue circles) and $t=3$ (red squares), while solid black lines stand for the analytical approximation obtained in \ref{app:Laplace-method}, which asymptotically converges to the rough estimation given by equation \eqref{eq:Approximation_NESS}. Vertical dashed lines at $|x|/t= \sqrt{4 D r_1}$ indicate the separation between the inner region, where the NESS has already been reached, and the outer region, where the transient behaviour is still observed.}
    \label{fig:RelaxationNESS}
\end{figure}

The comparison between simulations and the exhaustive analytical approach, obtained in \ref{app:Laplace-method}, in figure \ref{fig:RelaxationNESS} shows an excellent agreement.  We remark that, despite the model and the derivation of the large deviation function in equation \eqref{eq:Approximation_NESS} is considerably more involved than in the simpler case of resetting without refractory periods, the change of behaviour of the large deviation function corresponds to a second-order dynamical phase transition, where the second derivative of the Large Deviation Function is discontinuous at the matching point $\frac{|x-x_0|}{t} = \sqrt{4 D r_1}$ \cite{Majumdar15_Relaxation}. Therefore, the inclusion of the refractory periods does not affect this interesting feature. 

\section{First passage time with refractory periods}
\label{sec:MFPT}

In this section, we consider the first passage time problem to a target point $x_t$. In the absence of refractory period ($r_2 \to \infty$), it is known that there appears an optimal resetting rate $r_1^{\opt}$ that minimises the MFPT~\cite{Evans11_resetting,Evans11_optimal}. Here, we are interested in the analysis of the effect of the refractory period on the MFPT. On a physical basis, it is clear that the MFPT will increase as $r_2$ decreases, i.e. as the time spent at rest increases. Still, since the probability distributions of resetting events and refractory periods are independent, one might naively think that the optimal resetting rate $r_1^{\opt}$ would remain unaffected by $r_2$. The analysis below shows that this expectation is not fulfilled: in fact, the optimal resetting rate $r_1^{\opt}$ presents a non-trivial dependence on $r_2$.

\subsection{General formulation}

Including an absorbing boundary at $x_t$ in our system makes the former expressions invalid: normalisation to unity is not preserved anymore. For the sake of clarity, we do not explicitly add the parametric dependence on $x_t$, but it is important to keep in mind that all functions depend on it from now on, e.g. $p^{\free}(x,t|x_0)$ refers to the propagator in absence of resetting of a particle starting at $x_0$ with an absorbing boundary at $x_t$. For every propagation phase prior to a resetting event, we have to consider that the particle has not reached the target point $x_t$. Let  $Q^{\free}(x_0,t)$ denote the free survival probability, i.e. the probability of not being absorbed by the target for a time interval $t$  in the absence of resetting, provided that the particle started propagating from $x_0$ at $t=0$. 

The probability of finding the particle in $x$ at time $t$ fulfills the renewal equation
\begin{eqnarray}
    p(x,t|x_0)=&F(t) p^{\free}(x,t|x_0) \nonumber\\
    &+\int_0^t \d t_1 \, f(t_1)Q^{\free}(x_0,t_1) W(t-t_1) \delta(x-x_0) \nonumber\\
    &+\int_0^t \d t_1 \, f(t_1)Q^{\free}(x_0,t_1) \int_{t_1}^t \d \tau_1 \, w(\tau_1-t_1) \,p(x,t-\tau_1|x_0).
\end{eqnarray}
Its integral over $x$ gives us the renewal structure of the survival probability with resetting $Q_r(x_0,t)$,
\begin{eqnarray}
    \fl \qquad \quad Q_r(x_0,t)&=F(t)Q^{\free}(x_0,t) \nonumber\\
    &\quad +\int_0^t \d t_1 \, f(t_1)Q^{\free}(x_0,t_1) W(t-t_1) \nonumber\\
    &\quad +\int_0^t \d t_1 \, f(t_1)Q^{\free}(x_0,t_1) \int_{t_1}^t \d \tau_1 \, w(\tau_1-t_1) \,Q_r(x_0,t-\tau_1)
    \nonumber\\
    &=F(t)Q^{\free}(x_0,t)+\left[\left(f Q^{\free}\right)*w*Q_r\right](x_0,t)
    +\left[\left(f Q^{\free}\right)*W\right](x_0,t).
\end{eqnarray}
Note that now the resetting distribution is weighted with the survival probability when compared to the expressions obtained before.

To emphasise the power of our pathway formulation, we carry out an analysis completely similar to that in \Sref{section:ResettingPathways}. That is, we  expand the PDF in a series of terms $p_n^{\diff,\refr}$ corresponding to a given number of renewals $n$,
\begin{subequations}
\begin{align}
    p_n^{\diff}(x,t|x_0)&= \prod_{i=1}^n \left[
        \int_{\tau_{i-1}}^{t} \d t_i\,f(t_i-\tau_{i-1})Q^{\free}(x_0,t_i-\tau_{i-1})
        \int_{t_i}^{t} \d \tau_i\,w(\tau_i-t_i)
    \right] 
    \nonumber\\
    &\qquad \times F(t-\tau_n)p^{\free} (x,t-\tau_n |x_0) 
    \nonumber\\
    &=\left\{\left[ \left(f Q^{\free}\right)*w\right]^{*n} *\left(F p^{\free}\right)\right\} (x,t),
    \label{eq:pndiff_Absorbing} \\
    p_n^{\refr}(x,t|x_0)&=\prod_{i=1}^{n} \left[
         \int_{\tau_{i-1}}^{t} \d t_i\,f(t_i-\tau_{i-1})Q^{\free}(x_0,t_i-\tau_{i-1})
        \int_{t_i}^{t} \d \tau_i\,w(\tau_i-t_i)
    \right] \nonumber\\
    &\quad \times \int_{\tau_{n}}^t \d t_{n+1}\, f(t_{n+1}-\tau_{n})Q^{\free}(x_0,t_{n+1}-\tau_{n}) W(t-t_{n+1})\delta(x-x_0)
    \nonumber\\
    &=\left\{
        \left[ \left(f Q^{\free}\right)*w\right]^{*n}*\left(f Q^{\free}\right)*W
    \right\} (t)\delta(x-x_0).
    \label{eq:pnrefr_Absorbing}
\end{align}
\end{subequations}
With this approach, we get 
\begin{equation}
    \Tilde{Q_r}(x_0,s)=\frac{1}{1-\Tilde{f Q^{\free}}(x_0,s) \Tilde{w}(s)}
    \left[
    \Tilde{F Q^{\free}}(x_0,s)+\Tilde{f Q^{\free}}(x_0,s) \Tilde{W}(s)
    \right]. \label{eq:SurvivalProb_Resetting}
\end{equation}
Of course, this general result for arbitrary $f$ and $w$ is consistent with the results in the literature \cite{Evans19_refractory,Maso19_residence}.

\subsection{Poissonian resetting and refractory period}

Now we consider the same model as in subsection \ref{sec:Pois2}, the resetting time and refractory period distributions are both Poissonian and given by equations \eref{eq:Distribution_Resetting} and \eref{eq:Distribution_Refractory}. The propagator in the presence of the absorbing boundary is
\begin{equation}
    p^{\free}(x,t|x_0)=\frac{1}{\sqrt{4\pi D t}}
    \left\{
    \exp\left[-\frac{(x-x_0)^2}{4Dt}\right]-\exp\left[-\frac{(x+x_0-2 x_t)^2}{4Dt}\right]
    \right\},
    \end{equation}
which is valid for all $x_t\in \mathbb{R}$. The free survival probability and its Laplace transform are  
\begin{equation}
Q^{\free}(x_0,t)=
    \mbox{erf}\left(\frac{1}{2}\sqrt{\frac{\taud}{t}}\right), \quad 
    \Tilde{Q^{\free}}(x_0,s)=\frac{1-e^{-\sqrt{\taud s}}}{s},
\end{equation}
where we have defined the characteristic diffusion time between the initial position and the target as
\begin{equation}
  \taud=\frac{(x_t-x_0)^2}{D}\geq 0.
\end{equation}
Note that $\taud$, and thus $Q^{\free}$ and $\Tilde{Q^{\free}}$, only depends on the distance $|x_t-x_0|$ between the initial position and the target. As a consequence, equation \eref{eq:SurvivalProb_Resetting} becomes
\begin{eqnarray}
    \Tilde{Q_r}(x_0,s)&=(s+r_1+r_2) 
    \frac{\Tilde{Q^{\free}}(x_0,s+r_1)}{s+r_2-r_1 r_2 \Tilde{Q^{\free}}(x_0,s+r_1)} \label{eqn-43}\\
    &=(s+r_1+r_2)\frac{1-e^{-\sqrt{\taud (s+r_1)}}}{s(s+r_1+r_2)+r_1 r_2e^{-\sqrt{\taud (s+r_1)}}}. \label{eq:tilde-Qr-x0-s}
\end{eqnarray}

\subsubsection{Mean first passage time.}

Equation \eref{eq:tilde-Qr-x0-s} for $\Tilde{Q_r}(x_0,s)$ cannot be easily inverted to time domain, but it represents an excellent resource to compute some relevant physical properties in an exact way. Let us introduce the first passage density of the particle being absorbed by the target:
\begin{equation}
    f_{\text{FPT}}(t;\tau_d)=
    -\frac{\partial  Q_r(x_0,t)}{\partial t},
\end{equation}
where we have introduced explicitly in the notation the parametric dependence on $\tau_d$. Therefrom, we derive the MFPT as the mean absorbing time,
\begin{eqnarray}
    T(r_1,r_2;\taud)&=\int_0^\infty \d t\, t \, f_{\text{FPT}}(t;\tau_d)=
    \lim_{s \to 0} \Tilde{Q_r}(x_0,s),
\end{eqnarray}
which is thus easily computable from equation \eref{eq:tilde-Qr-x0-s},
\begin{eqnarray}
    T(r_1,r_2;\taud)&=\left(e^{\sqrt{\taud r_1}}-1\right)\left(\frac{1}{r_1}+\frac{1}{r_2}\right). \label{eq:MFPT}
\end{eqnarray}
Since we are interested in the dependence of the MFPT on the parameters controlling the typical duration of the reset events, $r_1$, and the refractory periods, $r_2$, we have introduced them explicitly in the notation.
As expected on a physical basis, two contributions appear in the MFPT coming from the two summands in the second parenthesis. The first one, which depends exclusively on $r_1$, corresponds to the instantaneous resetting without refractory period \cite{Evans11_resetting,Evans11_optimal}. The second one stems from  the refractory period that we have introduced after each resetting event. 
\begin{figure}
    \centering
    \includegraphics[width=0.8\textwidth]{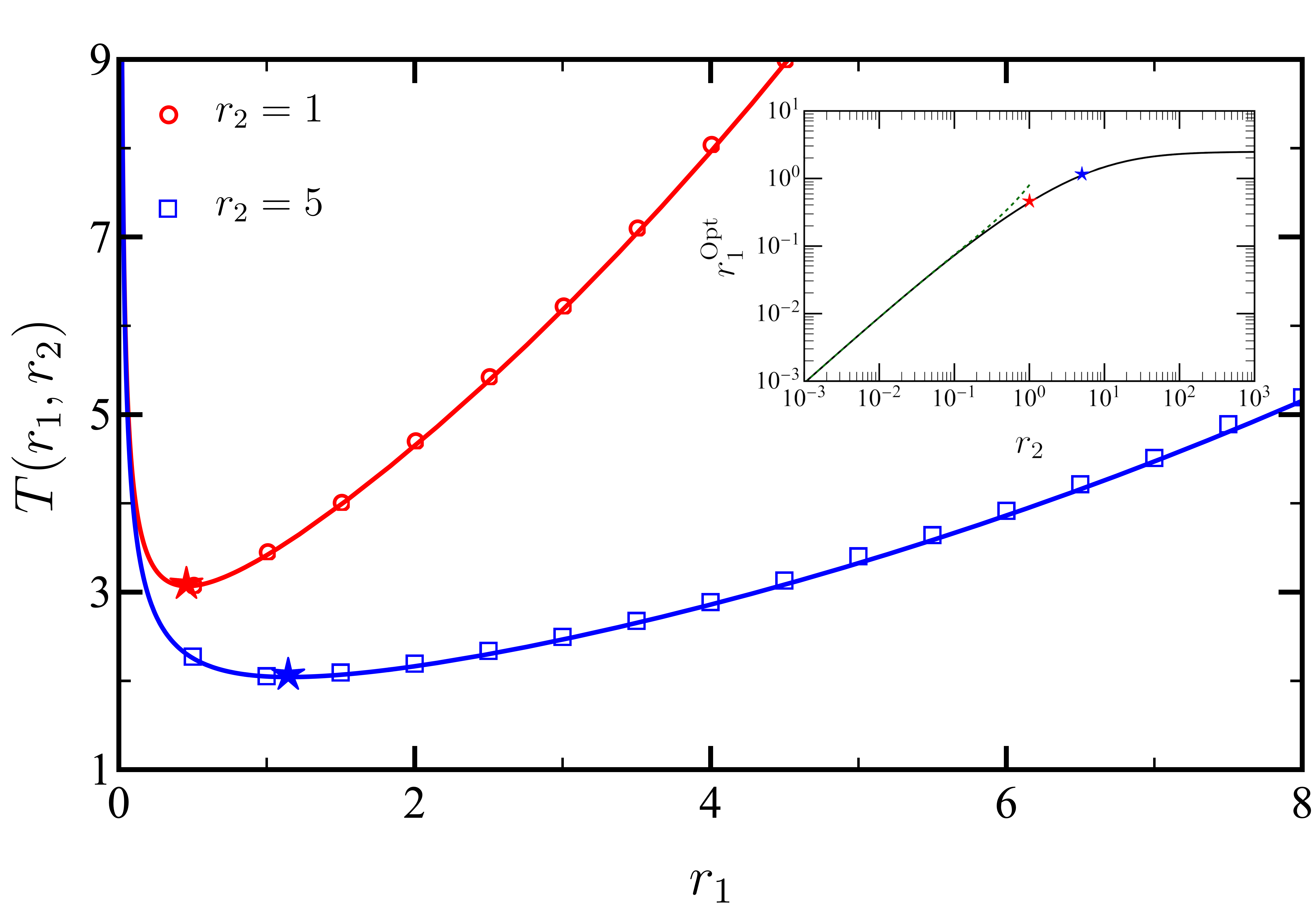}
    \caption{Mean first passage time $T(r_1,r_2)$ as a function of the resetting rate $r_1$ for fixed refractory rate $r_2$.  An excellent agreement between simulations (symbols) and theory (solid lines) is found. The optimal resetting rate $r_1^{\opt}$, represented by five-pointed stars, monotonically increases with $r_2$---in the limit $r_2\to \infty$, the value $r_1^{(0)}=\left(2-W(-2e^{-2})\right)^2= 2.53964$ is reached. Inset: Optimal resetting rate as a function of the refractory period rate.  We show the numerical solution of the implicit equation for $r_1^{\opt}$, as given by equation \eref{eq:DerivativeMFPT} (solid line), and the analytical approximation for small $r_2$, as given by equation \eref{eq:expansion_r1(r2)} (dashed line).}
    \label{fig:Tr1r2}
\end{figure}

Despite its simple functional form, equation \eref{eq:MFPT} exhibits atypical and rich behaviour. We are interested in how the MFPT changes as a function of $r_1$ and $r_2$---clearly, it monotonically increases with $\taud$, that is, it increases with the distance $|x_t-x_0|$ and the inverse of diffusion coefficient $D$. With this knowledge, we are able to write equation \eref{eq:MFPT} as a function which only depends on $r_1$ and $r_2$ introducing dimensionless parameters $r_1=r_1^*/\tau_d$, $r_2=r_2^*/\tau_d$ and $T=T^*\tau_d$. Therefore, the MFPT is
\begin{equation}
    T(r_1,r_2)=\left(e^{\sqrt{r_1}}-1\right)\left(\frac{1}{r_1}+\frac{1}{r_2}\right) \label{eq:MFPT_dimensionless},
\end{equation}
where we have dropped the asterisk in order not to clutter the formulae. This result was also obtained in an apparently different context~\cite{Mercado-Vasquez_2020}, when considering and intermittent V-shaped potential that was randomly switched on and off with different rates. Reasonably, in the limit of the stiffness of the potential going to infinity, the stochastic resetting with refractory periods is recovered.

\subsubsection{Optimal resetting rate \texorpdfstring{$r_1^{\opt}(r_2)$}.}

Let us first look for the minimum of $T(r_1,r_2)$ as a function of $r_2$ for fixed $r_1$. We clearly get that $r_2\to \infty$ for the best choice, reobtaining the MFPT in the absence of refractory period \cite{Evans11_resetting,Evans11_optimal},
\begin{equation}
\lim_{r_2 \to \infty} T(r_1,r_2)=\frac{e^{\sqrt{r_1}}-1}{r_1}.
\end{equation}
As a function of $r_1$, the minimum of $T(r_1,r_2\to\infty)$ is reached at $r_1=r_1^{(0)}=\left(2+W(-2e^{-2})\right)^2$, where $W(x)$ corresponds to the Lambert $W$ function, which is the inverse function of equation $f(x)=x e^x$, i.e. the solution of $W(x)\exp[W(x)]=x$~\cite{Evans11_resetting,Evans11_optimal}. 

Now we focus on looking for the minimum of $T(r_1,r_2)$ as a function of $r_1$ for fixed $r_2$. In figure \ref{fig:Tr1r2}, we show a couple of instances of $T(r_1,r_2)$ as a function of $r_1$ for $r_2 \in \{1,5\}$, finding an excellent agreement between theory and simulations. The figure highlights the existence of a certain optimal curve $r_1^{\opt}=r_1^{\opt}(r_2)$, which is obtained by solving
\begin{equation}
    0=\left.\frac{\partial T(r_1,r_2)}{\partial r_1}\right|_{r_1^{\optt}}
    =
    \left.
    \frac{2 r_2\left(1- e^{\sqrt{r_1}}\right)
    +\sqrt{r_1}e^{\sqrt{r_1}}\left(r_1+r_2\right)}
    {2 r_1^2 r_2}
    \right|_{r_1^{\optt}}. 
\end{equation}
Then, $r_1^{\opt}$ is implicitly given as a function of $r_{2}$ by
\begin{equation}
    2 r_2\left(1- e^{\sqrt{r_1^{\optt}}}\right)
    +\sqrt{r_1^{\opt}}e^{\sqrt{ r_1^{\optt}}}\left(r_1^{\opt}+r_2\right)=0.
    \label{eq:DerivativeMFPT}
\end{equation}

Now we analyse the limiting behaviour of $r_1^{\opt}$ for both $r_2\to\infty$ (no refractory period) and $r_2\to 0$ (infinite refractory period).  On the one hand, for $r_2 \to \infty$, we obtain that the limiting value $\lim_{r_2 \to \infty } r_1^{\opt}(r_2)$ is determined by
\begin{equation}
   \lim_{r_2 \to \infty} 2 \left[r_1^{\opt}(r_2) \right]^{-1/2}\left[ 1-e^{-\sqrt{ r_1^{\optt}(r_2)}}\right]= 1,
\end{equation}
i.e. $\lim_{r_2 \to \infty } r_1^{\opt}(r_2)=r_1^{(0)}$---which is consistent with the optimal resetting strategy without refractory period previously introduced. On the other hand, it is clear from equation \eref{eq:DerivativeMFPT} that $\lim_{r_2\to 0}r_1^{\opt}(r_2)=0$, which is logical from a physical point of view: for infinite refractory period, the best strategy for the MFPT is to avoid resetting. A dominant balance argument shows that $r_1^{\opt}\sim r_2$ in this limit.

In order to further investigate the dependence of $r_1^{\opt}$ on $r_2$ for long refractory periods (small $r_2$), it is handy to expand $r_1^{\opt}$ in a power series of  $\sqrt{r_2}$. 
Substituting this expansion into equation \eref{eq:DerivativeMFPT}, one gets after a little bit of algebra 
\begin{equation}
    r_1^{\opt}=r_2-r_2^{3/2}+\frac{5}{6}r_2^2+O\left(r_2^{5/2}\right). 
   \label{eq:expansion_r1(r2)}
\end{equation}
A comparison between the expansion \eref{eq:expansion_r1(r2)} and the numerical estimate for $r_1^{\opt}$ is shown in the inset of figure \ref{fig:Tr1r2}.

Our result shows that there appears a ``resonance" phenomenon, which optimises the MFPT---making it minimum---for a resetting rate that is linked with the refractory period rate. When the resetting point $x_0$ and the target $x_t$ are close, in the sense that $r_2 \taud\ll 1$, $r_1^{\opt} \simeq r_2$. 
As $r_2$ is increased, $r_1^{\opt}$ consequently increases but it asymptotically saturates for large enough values of $r_{2}$: for $r_2 \taud\gg 1$, the optimum MFPT asymptotically tends to  its limiting value $r_1^{(0)}$ corresponding to stochastic resetting without refractory periods.

\section{Conclusions}
\label{sec:concl}

In this work, we have carried out a thorough analysis of the effects of introducing a time cost to stochastic resets in a one-dimensional Brownian searcher. First, we have exploited a pathway formulation to derive general results. This puts forward an alternative, appealing from a physical point of view, methodology to address the study of intermittent dynamics.  Second, we have particularised the results for the relevant case of Poissonian resetting events and refractory periods. Therein, not only have we obtained the non-equilibrium stationary state, but also a detailed solution of the transient dynamics in the time domain. Finally, we have studied in depth the single-target search problem with refractory period. Specifically, we have investigated the optimal strategy for resetting, in the sense of minimising the mean first passage time to the target, finding that the optimal resetting rate depends on the typical duration of the refractory phase. 

From a physical perspective, the final result on the optimal resetting rate to find a target is especially interesting.  The dependence of the optimal resetting rate on the refractory period is somehow counterintuitive, since the duration of the time intervals between resetting events and the duration of the refractory phase are independent in our model. The optimal strategy entails a non-trivial, resonance-like behaviour in which the optimal resetting rate equals the inverse of the characteristic time of the refractory period. This phenomenon is reminiscent of resonant activation \cite{Doering92_resonant,Boguna98_resonant} where, when considering the escape problem in a two-well potential mediated by a fluctuating barrier, an optimal fluctuation rate that minimises the escape time emerges. Indeed, our resetting setup with refractory periods can be thought of as a fluctuating potential that switches between a totally confining potential trap and zero. Just recently, the potential connection between optimal resetting and resonant activation has started to be explored \cite{Capala21_dichotomous}, which provides an interesting perspective for further  research.   

\section*{Acknowledgments}
C.~A.~Plata acknowledges the funding received from European Union's Horizon Europe--Marie Sk\l{}odowska-Curie 2021 programme through the Postdoctoral Fellowship with Ref.~101065902 (ORION). G.~García-Valladares, C.~A.~Plata and A.~Prados acknowledge financial support from Grant PID2021-122588NB-I00 funded by MCIN/AEI/10.13039/501100011033/ and by ``ERDF A way of making Europe'', and also from Grant ProyExcel\_00796 funded by Junta de Andalucía's PAIDI 2020 programme. D.~Gupta acknowledges the Nordita fellowship program. Nordita is partially supported by Nordforsk.

\section*{Data availability}
The codes employed for generating the data that support the findings of this study, together with the Mathematica notebooks employed for producing the figures presented in the paper, are openly available in the~\href{https://github.com/fine-group-us}{GitHub page} of University of Sevilla's FINE research group.

\appendix

\section{Asymptotic analysis of the PDF for the propagation phase in the long-time limit} \label{app:Laplace-method}

In this appendix, we derive an approximate explicit expression for the different integrals appearing in equation \eref{eq:preLap}. For the sake of compactness, in the remaining of this appendix, we take $x_0=0$. The general result can be reobtained at the end with the substitution $x \to x-x_0$.

We start by focusing on the first integral term in equation \eref{eq:preLap},
\begin{eqnarray}
    I_1 = \int_0^1 \,\d\omega \,\omega^{-1/2}\,e^{-r_1 t \, [\omega + x^2 / (4 D \omega r_1 t^2)]} 
    = \int_0^1 \,\d\omega \,\omega^{-1/2}\,e^{-  \omega t^*- \frac{{x^*}^2}{\omega t^*}},
\end{eqnarray}
where we have introduced the following dimensionless variables $t^* = r_1 t$ and $x^*= x / \sqrt{4 D r_1^{-1}}$. In the following, we drop the asterisks for the sake of a clearer notation. For long times,  Laplace's method tells us that the integral is dominated by the maximum of the exponent, i.e. the minimum of $\phi_{1}(\omega)= \omega t + x^2/(\omega t)$ inside the integration interval $(0,1)$. Since the function $\phi_{1}(\omega)$ has a relative minimum at $\omega_0 = |x|/t$, how to estimate the integral depends on the value of $\omega_0$, specifically on whether $\omega_{0}$ is larger or smaller than unity. Then, it is handy to introduce the change of variable $\omega = \omega_0 \nu$,
\begin{equation}
    I_1 = \omega_0^{1/2}\int_0^{1/\omega_0} \,\d\nu \,\nu^{-1/2}\,e^{-\omega_0 t \, \psi (\nu)},
\end{equation}
where we we have introduced $\psi(\nu)=\nu+1/\nu$, which attains its relative minimum  at $\nu_0=1$. Now we asymptotically estimate $I_{1}$ in the long-time limit $t\gg 1$, with $\omega_0=\mathcal{O}(1)$, i.e. $x=\mathcal{O}(t)$. We must discriminate between different cases:
\begin{enumerate}
\item The relative minimum of $\psi(\nu)$ at $\nu=1$ lies inside the interval $(0,1/\omega_0)$, i.e. $\omega_0 < 1$ or $|x| < t$, and, in addition, it is far enough from the upper limit, in a sense that is clarified below.

The idea of Laplace's method is to expand $\psi(\nu)$ around the relative minimum at $\nu=1$,
 \begin{equation}
      \omega_{0}t\psi(\nu)\sim 2\omega_{0}t+\omega_{0}t(\nu-1)^{2},
 \end{equation}
 which leads to a Gaussian centred at $\nu=1$ and very small width, proportional to $(\omega_{0}t)^{-1/2}\ll 1$\footnote{Note that the following term in the Taylor series of $\omega_{0}t\psi(\nu)$ is $-\omega_{0}t(\nu-1)^{3}$, which is negligible against the retained quadratic term $\omega_{0}t(\nu-1)^{2}$ where the Gaussian contributes to the integral, i.e. for $\nu-1=O(\omega_{0}t)^{-1/2}$.}. Therefore, the dominant behaviour of the integral comes from a narrow interval around $\nu=1$. Here, we consider that $\omega_{0}$ is such that the whole Gaussian belongs in the integration interval $(0,1/\omega_{0})$, i.e. the integral is dominated by the contribution from $(1-\varepsilon,1+\varepsilon)$, with $\varepsilon\ll 1$, because we can choose $\varepsilon$ such that
\begin{equation}
    \label{eq:Appendix_ConditionA}
        \delta_{\inn} \equiv 1/\omega_0 - 1 >\varepsilon\gg (\omega_0 t)^{-1/2} .
\end{equation}
With this line of reasoning,
\begin{equation}
        I_1 \sim I^{(i)}_1 = \omega_0^{1/2}e^{-2\omega_0 t} \int_{1-\varepsilon}^{1+\varepsilon} \,\d\nu \,e^{-\omega_0 t \, 
        (\nu-1)^2}=\frac{1}{\sqrt{t}}e^{-2\omega_{0}t} \int_{-\varepsilon\sqrt{\omega_{0}t}}^{+\varepsilon\sqrt{\omega_{0}t}} \,\d z \,e^{-z^{2}}\sim\sqrt{\frac{\pi}{t}}e^{-2 |x|},
        \label{eq:Appendix_I_A}
\end{equation}
where condition \eref{eq:Appendix_ConditionA} allows for the integration limits in the last integral be extended to $\pm\infty$. Note that the obtained expression corresponds to the first case in equation \eref{eq:Approximation_NESS} if we reintroduce dimensions, i.e.\ the NESS behaviour \eref{eq:Ness_pprop}.
\item The relative minimum of $\psi(\nu)$ at $\nu=1$ lies outside the interval $(0,1/\omega_0)$, i.e. $\omega_0 > 1$ or $|x| > t$ and, in addition, it is far enough from the upper limit, in a sense that is also clarified below.
  
In this case, the minimum of $\psi(\nu)$ within the interval occurs at the upper limit $1/\omega_0$. Hence, Laplace's method tells us to expand $\psi(\nu)$ around $\nu=1/\omega_{0}$, $\psi(\nu)\simeq \omega_{0}+1/\omega_{0}+(1-\omega_{0}^{2})(\nu-1/\omega_{0})+\omega_{0}^{3}(\nu-1/\omega_{0})^{2}$ and restrict the integral to a narrow interval $(1/\omega_{0}-\epsilon,1/\omega_{0})$, with $\varepsilon\ll 1$. The quadratic term is negligible against the linear one if
 \begin{equation}\label{eq:Appendix_ConditionB1}
     \frac{\omega_{0}^{2}-1}{\omega_{0}^{3}}\gg \varepsilon.
 \end{equation}
 Assuming this ``far enough'' condition holds, we have 
 \begin{align}
        I_1 \sim I^{(ii)}_1 &= \omega_0\,e^{-t (1+\omega_0^2)}\int_{1/\omega_0-\varepsilon}^{1/\omega_0} \,\d\nu \,e^{\omega_0 t \, 
        \left(\omega_0^2-1\right)\left(\nu-1/\omega_0\right)}
                              \nonumber\\
                            &= \frac{e^{-t (1+\omega_0^2)}}{t \left(\omega_0^2-1\right)}
                              \int_{-\omega_{0}t \left(\omega_0^2-1\right)\varepsilon}^{0}
                              \,\d z \,e^{z}
        \sim {  \frac{e^{-t(1+x^2/t^2)}}{t\left(\frac{x^{2}}{t^{2}}-1\right)}, }
        \label{eq:apI1ii}
\end{align}
provided that the extension of the lower limit to $-\infty$ can be justified, i.e. we can choose $\varepsilon$ such that
\begin{equation}
    \omega_{0}t \left(\omega_0^2-1\right)\varepsilon \gg 1.
    \label{eq:Appendix_ConditionB2}
\end{equation}
  
Conditions \eref{eq:Appendix_ConditionB1} and \eref{eq:Appendix_ConditionB2} can be fulfilled without problems when $\omega_{0}-1=O(1)$, since they tell us that we have to choose $\varepsilon$ small but much larger than $(\omega_{0} t)^{-1}\ll 1$. As $\omega_{0}$ approaches unity, $1/\omega_{0}=1-\delta_{\out}$, with $\delta_{\out}\ll 1$, conditions \eref{eq:Appendix_ConditionB1} and \eref{eq:Appendix_ConditionB2} entail that
\begin{equation}\label{eq:Appendix_Condition_B3}
  \delta_{\out}\gg\varepsilon, \quad \omega_{0}t \delta_{\out}\varepsilon\gg 1 \implies \delta_{\out}\gg (\omega_{0}t)^{-1/2}. 
\end{equation}
This ``far enough'' condition makes sense: it is telling us that the separation of the upper limit from unity (the position of the relative minimum of $\psi$) must be much larger than the width of the Gaussian, analogously to condition \eref{eq:Appendix_ConditionA}.

Equation \eref{eq:apI1ii} has the same dominant contribution in the exponent that the non-resetting term $p^{\free}(x,t|x_0)$. However, the tails of the PDF are dominated by the non-resetting term, as expressed by equation \eref{eq:Approximation_NESS}, since the coefficient of equation \eref{eq:apI1ii}, $(x^2/t)^{-1}$, is subdominant compared to that in $p^{\free}$, $t^{-1/2}$, for $x/t=O(1)$.
\item The relative minimum of $\psi(\nu)$ at $\nu=1$ lies inside the interval $(0,1/\omega_0)$, i.e. $\omega_0 < 1$ or $|x| < t$,  but it is close to the upper limit,  $1/\omega_0 = 1+\delta_{\inn}$ with $\delta_{\inn} $ not fulfilling condition \eref{eq:Appendix_ConditionA}, i.e. $\delta_{\inn}=O(\omega_{0} t)^{-1/2}$. Herein, we have
\begin{align}
    I_1 \sim I^{(iii)}_1& = \omega_0^{1/2}e^{-2\omega_0 t} \int_{1-\varepsilon}^{1+\delta_{\inn}} \,\d\nu  \,e^{-\omega_0 t \, 
                        (\nu-1)^2}=\frac{1}{\sqrt{t}}e^{-2\omega_{0}t} \int_{-\varepsilon\sqrt{\omega_{0}t}}^{+\delta_{\inn}\sqrt{\omega_{0}t}} \,\d z \,e^{-z^{2}} \nonumber \\
  &\sim\frac{1}{2}\sqrt{\frac{\pi}{ t}}
                               \mbox{erfc}\left(\frac{|x|-t}{\sqrt{|x|}}\right)e^{-2 |x|},
\end{align}
where we have considered that $\varepsilon$ can always be choosen such that $\varepsilon \sqrt{\omega_0 t} \gg 1$. Note that this expression asymptotically converges to $I^{(i)}_1$ in equation \eref{eq:Appendix_I_A} when the limit  $\delta_{\inn} \sqrt{\omega_0 t} \gg 1$ is considered. Thus, this expression may be used to approximate $I_1$ for $\omega_{0}<1$, i.e. $|x|<t$, regardless of the value of $\delta_{\inn}$. 
\item The relative minimum of $\psi(\nu)$ at $\nu=1$ lies outside the interval $(0,1/\omega_0)$, i.e. $\omega_0 > 1$ or $|x| > t$ but it is close to the upper limit, i.e. $1/\omega_0 = 1-\delta_{\out}$ with $\delta_{\out} $ not fulfilling condition \eref{eq:Appendix_Condition_B3}, i.e. $\delta_{\out}=O(\omega_{0} t)^{-1/2}$.

Following the general idea of the Laplace method, we expand $\psi(\nu)$ around $1/\omega_0=1-\delta_{\out}$ in a narrow interval $(1-\delta_{\out}-\varepsilon, 1-\delta_{\out})$.  In (ii), $\varepsilon\ll\delta_{\out}$, but here we consider that $\varepsilon$ is at least $O(\delta_{\out})$. Therefore, we get
\begin{equation}
        \omega_{0}t\psi(\nu)\simeq \omega_{0}t \left[2 +\delta_{\out}^{2}+ 2\delta_{\out}(\nu - 1 +\delta_{\out}) + (\nu - 1 + \delta_{\out})^2\right],
\end{equation}
neglecting $O(\omega_{0}t\delta_{\out}^{3})$, $O(\omega_{0}t\delta_{\out}^{2}\varepsilon)$, $O(\omega_{0}t\delta_{\out}\varepsilon^{2})$, and $O(\omega_{0}t\varepsilon^{3})$ terms.\footnote{Recalling that $\delta_{\out}=O(\omega_{0} t)^{-1/2}$, we have, on the one hand, $O(\omega_{0}t\delta_{\out}^{3})=O(\omega_{0}t)^{-1/2}\ll 1$, $O(\omega_{0}t\delta_{\out}^{2}\varepsilon)=O(\varepsilon)\ll 1$. On the other hand, both $O(\omega_{0}t\delta_{\out}\varepsilon^{2})=O((\omega_{0}t)^{1/2}\varepsilon^{2})$ and $O(\omega_{0}t\varepsilon^{3})$ must be much smaller than unity, so $\varepsilon\ll(\omega_{0}t)^{-1/3}$ .} Introducing this expansion, we obtain
\begin{equation}
   I_{1}\sim \omega_{0}\, e^{-(2+\delta_{\out}^{2})\omega_{0}t}\int_{1-\delta_{\out}-\varepsilon}^{1-\delta_{\out}} \, d\nu\, e^{-\omega_{0}t \left[2\delta_{\out}(\nu - 1 +\delta_{\out}) + (\nu - 1 + \delta_{\out})^2\right]}.
 \end{equation}
The change of variables $2\delta_{\out}\omega_0 t(\nu - 1 +\delta_{\out})=z$ allows us to write
    \begin{equation}
        I_1 \sim  \frac{1}{2\delta_{\out}t}\, e^{-(2+\delta_{\out}^{2})\omega_{0}t}\int_{-2\delta_{\out}\omega_0 t \varepsilon}^{0} \,\d z \,
        e^{ z - z^2/(4 \delta_{\out}^2 \omega_0 t)}.
    \end{equation}
Now we choose $\varepsilon$ such that $\delta_{\out}\omega_0 t \varepsilon\gg 1$, which makes it possible to extend the lower limit of the integral to $-\infty$, similarly to the other cases we have analysed.\footnote{Taking into account, once again, that $\delta_{\out}=O(\omega_{0}t)^{-1/2}$, this implies that $\varepsilon\gg (\omega_{0}t)^{-1/2}$. So our small parameter $\varepsilon$ must verify $(\omega_{0}t)^{-1/3}\gg\varepsilon\gg (\omega_{0}t)^{-1/2}$, e.g. we may take $\varepsilon\propto (\omega_{0}t)^{-2/5}$.} Finally, the explicit approximation we get is
    \begin{align}
      I_1 \sim I^{(iv)}_1 &=\frac{1}{2\delta_{\out}t}\, e^{-(2+\delta_{\out}^{2})\omega_{0}t}\int_{-\infty}^{0} \,\d z \,
                            e^{ z - z^2/(4 \delta_{\out}^2 \omega_0 t)}\nonumber \\
      &=\frac{1}{2\delta_{\out}t}\, e^{-(2+\delta_{\out}^{2})\omega_{0}t}\,\sqrt{\pi\delta_{\out}^{2}\omega_{0}t}\, e^{\delta_{\out}^{2}\omega_{0}t}\mbox{erfc}\left(\sqrt{\delta_{\out}^{2}\omega_{0}t}\right) \nonumber \\
                          &=\frac{\sqrt{\pi |x|}}{2t}\,e^{-2 |x|}\mbox{erfc}\left(\frac{|x|-t}{\sqrt{|x|}}\right).
    \end{align}
  \end{enumerate}

The other integral term involved in equation \eref{eq:preLap} has a simpler analysis. We employ again dimensionless variables, but now those stemming from the natural units evidenced by $\Phi_2$, i.e. $x^* = x/\sqrt{4 D r_2^{-1}}$ and $t^*=r_2 t$---and drop the asterisks once more. The exponent $\Phi_2(\omega;y)$ is a monotonically increasing function of $\omega$. Thus, the local maximum of the exponent within the integration interval occurs for all cases at the upper limit. Similarly to the case (ii) for $I_1$, direct application of Laplace's method gives
\begin{equation}
    I_2 = \int_0^1 \,\d\omega \,\omega^{-1/2}\,e^{  \omega t - \frac{x^2}{\omega t}}  \sim \frac{e^{t (1 -x^2/t^2)}}{t(1+x^2/t^2)} = \left(x^2 / t + t \right)^{-1} e^{t(1-x^2/t^2)}.
\end{equation}

In figure \ref{fig:RelaxationNESS}, the evaluation of the theoretical prediction \eref{eq:preLap} is performed by computing the approximated expressions in this appendix for $I_1$ and $I_2$.\footnote{We recall that those expressions depend on dimensionless magnitudes that involve $r_1$ and $r_2$, respectively.} For the inner region, $I_1^{(iii)}$ is used, while $I_1^{(iv)}$ and $I_1^{(ii)}$  are used for the outer region. The change between $I_1^{(iv)}$ and $I_1^{(ii)}$ is made at the crossing point $x_{\mscriptsize{\mbox{cross}}}>0$ where $I_1^{(iv)} = I_1^{(ii)}$, which has been numerically obtained.
Hence, the plotted line stands for:
\begin{equation}
    \fl
    p^{\diff} (x,t|x_0)  \sim
    \left\{ 
    \begin{aligned}        
        p^{\free}(x,t|x_0)+I_1^{(iii)}-I_2&, \quad |x|<t,\\
        p^{\free}(x,t|x_0)+I_1^{(iv)}-I_2&, \quad t<|x|<x_{\mscriptsize{\mbox{cross}}},\\
         p^{\free}(x,t|x_0)+I_1^{(ii)}-I_2&, \quad x_{\mscriptsize{\mbox{cross}}}<|x|.
    \end{aligned} \right.
\end{equation}

\section*{References}
\bibliographystyle{iopart-num}
\bibliography{14_February_Resetting}

\end{document}